\documentclass{singlecol-new}

\usepackage{stfloats}
\usepackage[numbers]{natbib}
\usepackage{mathrsfs}
\usepackage{lineno,hyperref}
\usepackage{amsmath}
\usepackage{amssymb}
\usepackage{amstext}
\usepackage{algorithm}
\usepackage{algorithmic}
\usepackage{graphicx}

\theoremstyle{TH}{

}

\theoremstyle{THrm}{

}

\theoremstyle{THhit}{

}

\makeatletter
\def\theequation{\arabic{equation}}

\JOURNALNAME{\TEN{\it Int. J. Autonomic Computing,
Vol. \theVOL, No. \theISSUE, \thePUBYEAR\hfill\thepage}}%
\makeatother

\begin{document}%

\setcounter{page}{1}

\LRH{M. Genkin et~al.}

\RRH{Autonomic Architecture for Big Data Performance
Optimization}

\VOL{x}

\ISSUE{x}

\PUBYEAR{202X}

\BottomCatch

\CLline

\subtitle{}

\title{Autonomic Architecture for Big Data Performance
Optimization}

\authorA{Mikhail Genkin}
\affA{School of Computer Science,\\ Carleton University,\\ Ottawa, ON, Canada \\
E-mail: michael.genkin@carleton.ca}
\authorB{Frank Dehne}
\affB{School of Computer Science,\\ Carleton University,\\ Ottawa, ON, Canada \\}

\authorC{Anousheh Shahmirza}
\affC{School of Computer Science,\\ Carleton University,\\ Ottawa, ON, Canada \\}

\authorD{Pablo Navarro}
\affD{School of Computer Science,\\ Carleton University,\\ Ottawa, ON, Canada \\}

\authorE{Siyu Zhou}
\affE{School of Computer Science,\\ Carleton University,\\ Ottawa, ON, Canada \\}

\begin{abstract}
The big data software stack based on Apache Spark and Hadoop has become mission critical in many enterprises. Performance of Spark and Hadoop jobs depends on a large number of configuration settings. Manual tuning is expensive and brittle. There have been prior efforts to develop on-line and off-line automatic tuning approaches to make the big data stack less dependent on manual tuning. These, however, demonstrated only modest performance improvements with very simple, single-user workloads on small data sets. This paper presents KERMIT - the autonomic architecture for big data capable of automatically tuning Apache Spark and Hadoop on-line, and achieving performance results 30\% faster than rule-of-thumb tuning by a human administrator and up to 92\% as fast as the fastest possible tuning established by performing an exhaustive search of the tuning parameter space. KERMIT can detect important workload changes with up to 99\% accuracy, and predict future workload types with up to 96\% accuracy. It is capable of identifying and classifying complex multi-user workloads without being explicitly trained on examples of these workloads. It does not rely on the past workload history to predict the future workload classes and their associated performance. KERMIT can identify and learn new workload classes, and adapt to workload drift, without human intervention.
\end{abstract}

\KEYWORD{autonomic computing; big data; machine learning;  high-performance computing.}

\REF{to this paper should be made as follows: Mikhail Genkin et al. (2023) `Autonomic Architecture for Big Data Performance Optimization', {\it International Journal of Autonomic Computing}, Vol. x, No. x, pp.xxx\textendash xxx.}

\begin{bio}
Mikhail Genkin received his Ph.D. in computer science at Carleton University in Ottawa, Canada. He is currently a Postdoctoral Fellow at the Toronto Metropolitan University in Toronto, Canada. Hi research focuses on applying big data, machine learning, and artificial intelligence towards designing autonomic software and systems.

Frank Dehne is a Professor at Carleton University in Ottawa, Canada. His research program focuses on building high performance computing systems for solving data intensive and/or computationally hard problems in business data analytics, biomedical data analytics and engineering data analytics. Current research projects include: (i) high performance data analytics for business intelligence, in particular real-time online analytical processing for multi-core and cloud architectures; (ii) high performance biomedical computing, in particular protein data analytics for protein-protein interaction prediction and drug design; and (iii) high performance data science platforms, in particular auto-tuning Hadoop and Spark.

Anousheh Shahmirza, Pablo Navarro, and Siyu Zhou are M.Sc. students at the Carleton Univesity School of Computer Science in Ottawa, Canada. Their research focuses on big data analytics and machine learning.

\end{bio}

\maketitle

\section{Introduction} \label{intro}

\subsection{Big Data Performance and Artificial Intelligence} \label{ingro-bd-ai}

Big data analytics have become mission critical in many enterprises.  Big data is used extensively in baking and securities, communications, media and entertainment, health care, education, manufacturing, govenment, insurance and retail industries. The use of big data has become so pervasive that today many enterprises not only augment their more traditional Relational Database Management Systems (RDBMS) with big data technologies, but have completely rebased their analytic processing around the big data technologies.

Big data processing presents unique challenges due to the sheer volume of the data that needs to be analyzed and to the fact that these data are typically loosely structured. While the functional aspects of these important big data frameworks are very important, so is performance. In many cases analytic jobs must complete their processing within a specified time duration in order for the results to be useful. Some big data jobs must complete their processing in sub-seconds. Recommender systems are one exmple of systems that need to execute big data very quickly in order to maintain an acceptable end-user experience. Other jobs, especially those that aim to analyze large volumes of historical data to enable accurate projections of future events, may take hours, days, or even weeks to complete.  

Today Apache Spark and Apache Hadoop form the foundation of the big data software stack. Other analytic technologies, such as Apache Hive, Apache Hbase, and Apache Tez, leverage these frameworks to implement their functionality. The performance of Spark and Hadoop jobs depends on a very large number of configuration settings. Manual tuning is expensive and brittle. Ideal combinations of tuning parameters have to be established experimentally. Each experiment often requires many hours to run due to the large volume of data involved. The introduction on new jobs, or changes in the nature or volume of the data, or the number of user executing jobs concurrently on the system oftem require the experiments to be repeated. A system capable of optimizing big data performance based on workload charactistics would be clearly beneficial, and would help greatly reduce operating costs of big data systems and improve the end user experince.

Recent advances in Artificial Intelligence (AI) and Machine Learning (ML) enabled new approaches to autonomic system design. Supervised ML algorithms can be used to classify objects and estimate their future quantifiable behavioral characteristics, such as speed and direction for example. Advanced supervised ML techniques, such as Zero-Shot Learning (ZSL), one-shot learning, and few-shot learning are now available, and can help reduce the effort associated with training data set construction. Unsupervised ML algorithms can be used effectively to discover patterns in data, thereby discovering previously unseen classes. Machine learning pipelines can be constructed to automate labeling and training of supervised ML algorithms. And so, the overarching research question behind this investigation is: "Can we combine ML techiques to construct a coherent architecture capable of optimizing big data performacne without human intervention?"

\section{Problem}

To date there has been a significant amount of research focusing on autonomic computing. The majority of relevant research has focused on the more traditional, 'small data' RDBMS space \cite{elnaffar2009psychicscheptic} \cite{holtzeritter2008ngram} \cite{linma2018selfdrivingdb}. Some research on this area also comes from network, cloud \cite{huangetal2017perfpredict} \cite{leietal2014hotspot}, and web \cite{cherkasova2014perfanomaly}, but there has been no research to date focusing on big data specifically.

Supervised learning algorithms investigated in \cite{genkin2019workloadclassification} and \cite{genkin2020workloadprediction} require explicit labelling of each workload and workload transition type. The research investigation in \cite{genkin2020zslmultiuser} would significantly reduce the required labelling and training effort, but would not eliminate it completely.

We don't want to replace the manual tuning problem with a potentially equally expensive labelling and training problem. We need a solution that can automate labelling and training functions and minimize human intervention.

\section{Limitations of Previous Approaches}

This is the first study that focuses on autonomic performance optimization for big data workloads. Previous works focusing on cloud and traditional small data systems have the following limitations:

\begin{enumerate}
\item Coarse view of workload, such as DSS vs. OLTP, makes it impossible to optimize job by job. Big data jobs can have very different optimal configuration parameters.
\item Linear regression models typically used to predict workload characteristics perform poorly with abrupt workload transition common in the big data space.
\item Most methods depend on past workload history to predict future workload characteristics. Coupled with coarse view of workload, this makes them ineffective in situations where workload characteristics change frequently.
\item None of the previous works include the ability to anticipate new, previously unseen, workloads.
\end{enumerate}

Most previous works focus on one of the aspects of autonomic computing but don't describe a complete architecture that implements an autonomic feedback loop.

\section{Contribution}

This work presents the first autonomic architecture specifically designed for autonomic optimization of big data workloads. This architecture implements the feedback loop based on pervasive implementation of machine learning algorithms. The Knowledge Extraction Resource Management Interface (KERMIT) architecture is able to interact with a wide variety of analytic frameworks and applications regardless of their internal architecture and resource usage pattern.

KERMIT can:

\begin{enumerate}
\item Learn new workloads and their characteristics without human intervention.
\item Anticipate the appearance of new, unseen, workload classes.
\item Detect changes in workload characteristics and classify workloads and workload transitions in real-time.
\item Predict which workload types and transitions are likely to occur in the future, and when.
\item Minimize parameter search overhead.
\end{enumerate}

Sections below will discuss previous work on autonomic architectures, and then go into a detailed description of the KERMIT architecture.

\section{Previous Work}

\begin{figure}[!t]
\centering
\includegraphics[width=\textwidth]{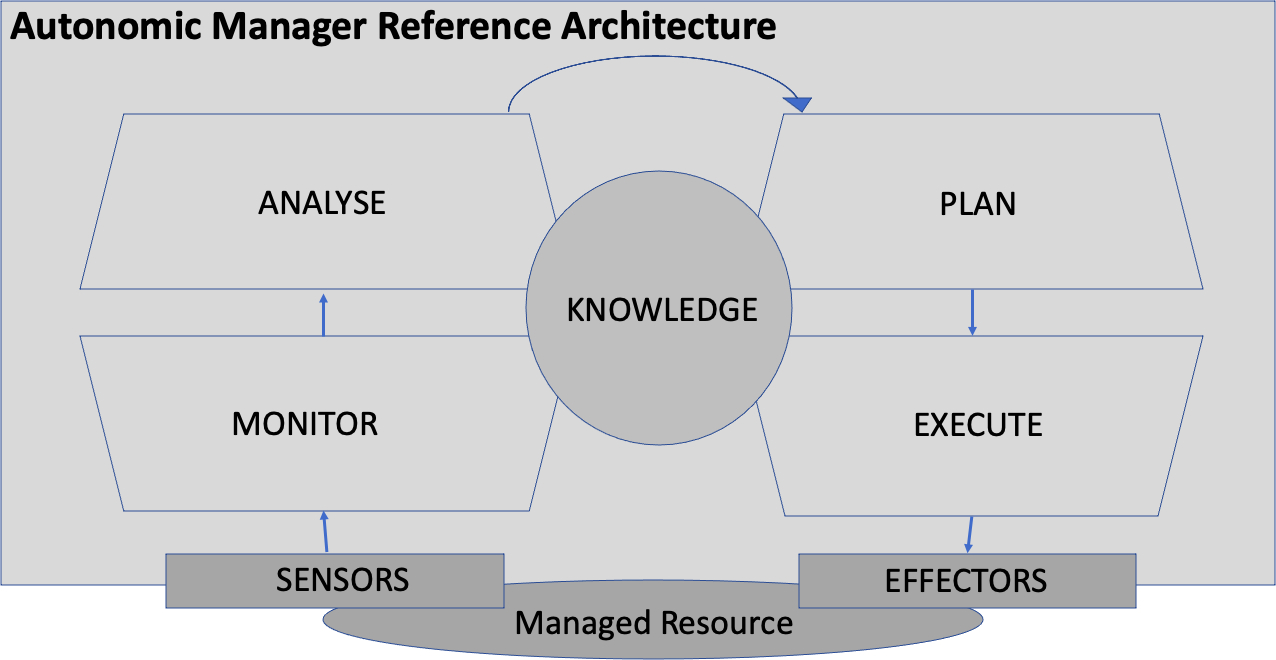}
\caption{The MAPE-K reference architecture for autonomic systems proposed by IBM. This figure was constructed based on the reference architecture description in \cite{mapekIBM}.}
\label{mape-k-ref-arch}
\end{figure}

The field of autonomic computing was introduced by IBM in 2001 with the goal of creating computer systems capable of self-management. In 2005 IBM published it's reference architecture for autonomic systems \cite{mapekIBM}. This reference architecture, referred to as MAPE-K, is shown in Figure ~\ref{mape-k-ref-arch}. The acronym MAPE-K stands for Monitor(M), Analyze(A), Plan(P), Execute(E), and Knowledge(K). 

Since then, there has been considerable research focusing on developing autonomic systems, but this paper is the first study focusing specifically on autonomic workload optimization for big data. This section summarizes the most relevant studies from the other problem domains. Works focusing on the cloud space and networking are discussed first, followed by more relevant works focusing on small data autonomic databases. 

Movahedi et al. \cite{autonomicNetworkSurvey} present a survey of autonomic network architectures (ANM). The authors classify ANM architectures into hierarchical and flat types. The authors noted that only one of the reviewed architectures - the Cognitive Network Architecture (CNA) - made use of learning techniques. The CNA architecture includes a cognitive plane responsible for data analysis and the decision-making process. They highlighted that the use of learning mechanisms could significantly improve the performance of policy-based adaptation schemes towards finding the optimal solution \cite{autonomicNetworkSurvey}.

Carrera et al. \cite{autonomicPlacement} present a study on autonomic placement of batch and transactional workloads. The authors present a technique that enables the existing middleware to fairly manage mixed workloads comprised of both batch analytic jobs and transactional applications. The authors define a simple objective function to measure the difference between the actual response-time and the response-time goal for interactive applications. Their architecture defines the placement control loop and an application placement controller component that periodically inspects the system to determine if placement changes are needed in response to the changing workload. The period of the control loop is configurable. The placement algorithm uses a mathematical model to estimate application performance relative to a given CPU allocation. Their method extrapolates the applications' performance over the duration of the current control cycle and subsequent cycles. The authors claim that their technique improves mixed workload performance while providing service differentiation based on high-level performance goals.

This approach would have limited applicability in the big data space. This is because the mathematical model used to estimate future performance is linear and would not be able to predict the very abrupt workload transitions that big data jobs present, such as the map-to-reduce transition that results in a major change in workload characteristics. Their method does not have any provision for learning. The mathematical model needs to be executed every time, even if similar workload transitions recur.

Gergin et al. \cite{gerginDecentralizedAutonomic} describe a decentralized autonomic architecture for performance control in the cloud. Their architecture utilizes feedback loops. It uses a series of autonomic controllers to monitor virtual machine utilization under a Web OLTP-type workload and provisions new virtual machines as needed to achieve SLA objectives. Each controller independently regulates a tier of the application and implements the proportional, integrative, and derivative control laws. The mathematical model underlying each controller uses linear component to extrapolate near-term performance. This approach, as discussed above, would not work well for big data applications because they tend to produce very abrupt workload transitions. Their architecture does not include a learning mechanism.

In their recent paper Nouri et al. \cite{nouriReinforcementCloud} focus on a cloud Infrastructure-as-a-Service (IaaS) use case. The authors describe a distributed architecture that aims to maximize performance of a large number of applications deployed on many servers. Each server is a virtual machine. Their view is that a centralized controller would become too complex because it would have to monitor a large number of applications on many server. It may not be able to respond in time when presented with a rapid change in load. A centralized controller would also become a single point of failure in the system.

Their architecture involves deploying an agent on each server. Each agent is responsible for monitoring the application performance on that server. The agents share a common knowledge base. Each agent has application and system monitoring components which feed information to a learning core. The architecture monitors the application response-time statistics and system resource utilization values for the server. 

The learning core is based on reinforcement learning. It maps a moving average of CPU utilization values to a set of states for the server. It then uses a utility function that converts that state into a scalar reward value that is used by reinforcement learning model to select from a set of actions for each state. The states actions and reward values are stored in the knowledge base on each server.  

Nouri et al. \cite{nouriReinforcementCloud} architecture is only of limited applicability to big data resource management because they consider a scenario whereby multiple applications running on a single server share a pool of resources. Thus exclusive resource allocation to the application is not possible. Containerized big data applications, on the other hand, rely on exclusive allocation of resources.

The architecture described in this work does not have any notion of searching the parameter space to optimize the application performance. Instead the controller can execute a limited set of actions to scale the number of application instances and/or servers up or down. Another aspect that limits this architectures applicability to KERMIT is the fact that this architecture uses linear regression to model system performance. Big data workloads present many abrupt and highly non-linear workload transitions. Nouri et al. architecture is reactive in nature. It does not predict future workload characteristics, and does not anticipate new workload classes.

In recent survey Raza et al. \cite{autonomicLargeRepSurvey} works focusing on autonomic performance tuning in large scale data repositories. This survey explicitly excludes works for focusing on big data. It focuses on the more traditional RDBMS-based data warehouses and on DSS and OLTP workloads. The authors organize research into several categories, including workload classification, performance prediction, and self-adaptation. Most of the surveyed papers focus on one of these aspects and only  a few combine them into an architecture that implements the full autonomic cycle. The most relevant ones are discussed in the paragraphs below.

Lin Ma et al. \cite{linma2018selfdrivingdb} describe QueryBot5000 (QB5000) - a system for query-based workload forecasting that can be used to implement autonomic qualities for RDBMS such as MySQL and PostgreSQL. Although QB5000 was able to forecast future workload characteristics, the authors do not explicitly describe how an autonomic loop could be implemented. 

Elnaffar and Martin \cite{elnaffar2009psychicscheptic} proposed a framework for predicting shifts in DBMS workload from predominantly OLTP-type to DSS-type. Their framework, called the Psychic-Skeptic Prediction (PSP) framework, included a mechanism for updating the prediction model in those situations where differences between the predicted and the actual workload characteristics were observed - thus implementing the autonomic feedback loop. The main limitation of this approach is that the PSP relied heavily on past workload cycles to determine whether the model needs to be updated.

In summary, most previous works focus on one of the aspects of autonomic computing but don't describe a complete architecture that implements an autonomic feedback loop. Most of the described methods tend to be reactive in nature. They don't have the ability to classify workloads or anticipate workload changes. Most methods depend on past workload history to predict future workload characteristics. This makes them less effective in situations where workload characteristics change frequently, or do not follow a regular pattern. None of the previous works include the ability to anticipate new, previously unseen, workloads.

Furthermore, those works that do implement the ability to anticipate workload changes have a very coarse view of workload, such as DSS vs. OLTP, making it impossible to optimize job by job. Big data jobs can have very different optimal configuration parameters. Linear regression models typically used to predict workload characteristics perform poorly with abrupt workload transition common in the big data space. Our architecture aims to address these shortcomings.

\section{Autonomic Architecture for Big Data Workload Optimization }

Before we delve into the details of the KERMIT architecture it is important to establish a clear description of the key concepts and terms used throughout this work. Once this is done we describe the key autonomic architecture principles, and finally we discuss, in detail, the design techniques used to implement them in KERMIT. 

\subsection{Key Concepts and Terminology}

The following key concept definitions underpin the KERMIT autonomic architecture:

\begin{enumerate}
 \item \textit{Workload}. Different researchers define the term \textit{workload} differently. In this work we use the definition presented during our earlier investigation \cite{genkin2020workloadprediction}. Consistent with our previous definition, this term refers to a multi-variate time-series of observation windows. Each observation window does not show statistically meaningful differences with neighbouring observation windows. Thus, workloads are uniquely idetifiable periods of steady-state processing. The concept of workload is represented with the symbol $\Omega$. 
 \item \textit{Workload transition}. A workload transition, like a workload, is a multi-variate time series of observation windows. Unlike a workload, it represents a period of non-steady-state processing. Each observation window of the workload transition will show statistically meaningful changes relative to the neighbouring windows.
 \item \textit{Workload drift}. Workload drift refers to changes, whether systematic or random, that can occur to a workload over time. This term is defined in \cite{genkin2020workloadprediction}. Sections below will further elaborate on this important concept.
\end{enumerate} 

\begin{figure}[!t]
\centering
\includegraphics[width=\textwidth]{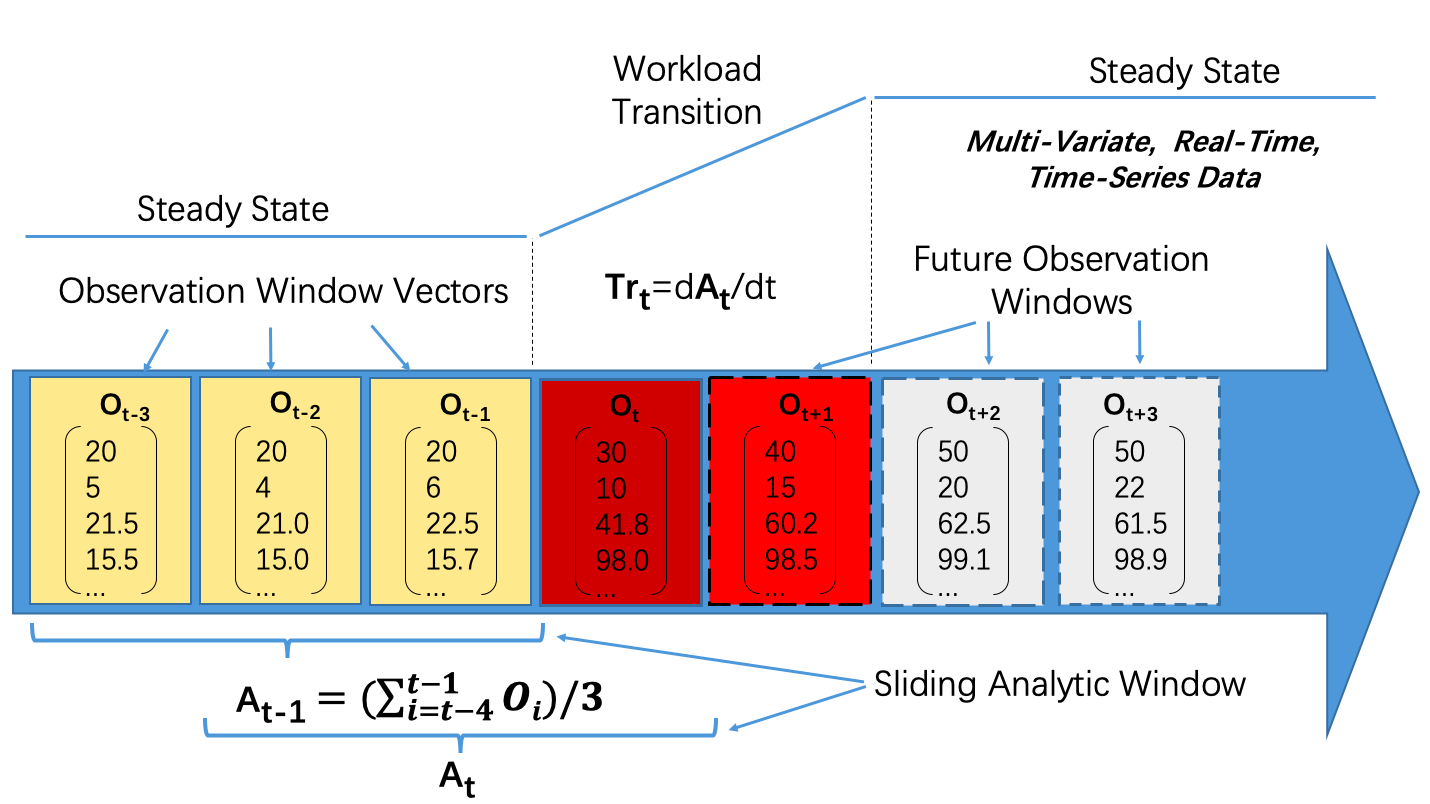}
\caption{Workloads are periods of steady state processing connected by workload transitions \cite{genkin2020workloadprediction}.}
\label{theory}
\end{figure}

The broader set of processing performed by the system can be described as a sequence of workloads connected together by workload transitions. This is shown in Figure ~\ref{theory} \cite{genkin2020workloadprediction}.

This view of workload is more granular than that generally used by other researchers (for example in \cite{holtzeritter2008ngram}, \cite{elnaffar2009psychicscheptic}). Nevertheless, it allows for a more systematic and automated treatment of workload analysis and optimization. 

\subsection{Autonomic Architecture Principles}

The MAPE-K architecture, shown in Figure ~\ref{mape-k-ref-arch} implements a feedback loop designed to achieve the key properties essential for an autonomic systems, as defined by IBM. These properties were defined as:

\begin{itemize}
\item \textit{Self-configuration}. An autonomic system must be able to configure, or re-configure, itself without requiring human intervention.
\item \textit{Self-healing}. In the event of a fault, an autonomic system must be able to find a way to mitigate the fault without human intervention.
\item \textit{Self-optimization}. An autonomic system must be able to optimize its performance within specified goals without human intervention.
\item \textit{Self-protection}. An autonomic system must be able to defend itself against misuse without human intervention.
\end{itemize}

Within the framework of the MAPE-K reference architecture (see Figure ~\ref{mape-k-ref-arch}) the Autonomic Manager uses Sensors to Monitor a Managed Resource. Monitoring generates Knowledge that pertains to the Managed Resource. This knowledge is Analysed and used to Plan the system's response to changes. During the Execute phase the Autonomic Manager uses the Effectors to affect the  Managed Resource in accordance with the Plan.

The research presented in this paper focuses on achieving the self-optimization characteristic. By extension some aspects of the self-configuration, self-healing, and self-protection characteristics are drawn in as well. This is due to the fact that in order to change the performance characteristics of the big data stack changes to the configuration are required. 

The self-healing and self-protection properties were not directly the subject of this investigation. Nevertheless, it can be said that the KERMIT architecture partially addresses those properties as well. For example, failure of one or more nodes in the cluster can present itself as the appearance of new workload types because the loss of the CPU and memory resources will alter the observed feature vector of the observation windows. The KERMIT architecture, as described below, will be able to react to this and find a new optimum. This type of response can-be characterised as partial self-healing. The self-healing is partial in this case because KERMIT does not address bring replacement nodes on-line. Similarly, KERMIT's ability to respond to new workload types can be viewed as partial protection form a sophisticated type of DoS/DDos attack targeting back end systems. 

Sections below describe how the KERMIT architecture layers map onto the MAPE-K reference architecture, and describe key components and algorithms in detail.

\subsection{The KERMIT Architecture}

\begin{figure}[!t]
\centering
\includegraphics[width=\textwidth]{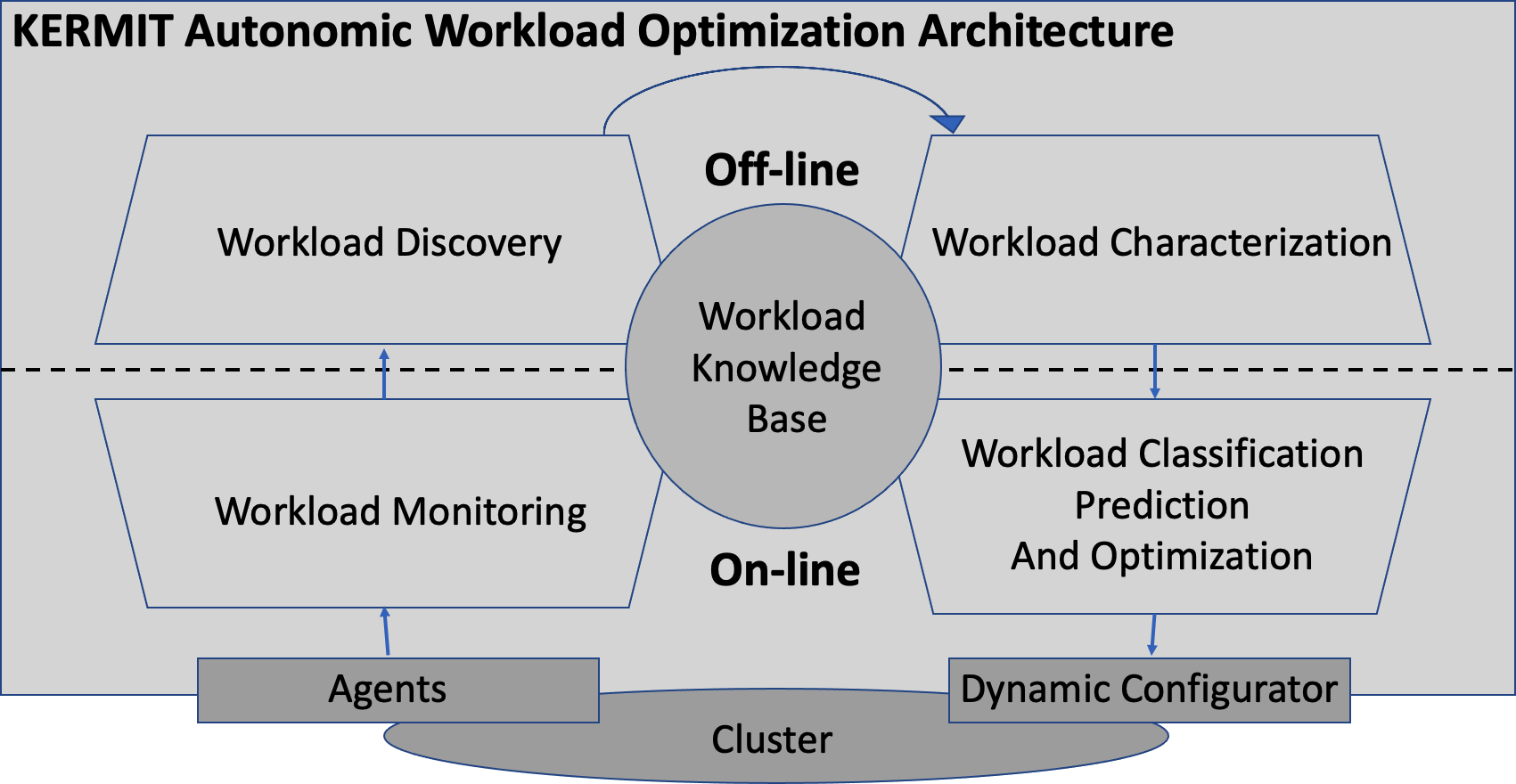}
\caption{Mapping the KERMIT sub-systems and components onto the MAPE-K reference architecture.}
\label{KERMIT-ref-arch}
\end{figure}

Figure ~\ref{KERMIT-ref-arch} shows the mapping of the key KERMIT sub-systems and components onto the MAPE-K reference architecture. The KERMIT architecture is broadly divided into two subsystems: 1 - On-line; 2 - Off-Line.

The On-line sub-system operates in real-time, and includes the Workload Monitoring sub-system, the Agents sub-system, the Dynamic Configurator sub-system, and the Workload Classification, Prediction and Optimization sub-system. The Off-line sub-system operates asynchronously in batch mode, and includes the Workload Discovery and the Workload Characterization sub-systems. All KERMIT sub-systems both update, and read from the Workload Knowledge Base sub-system. 

\begin{figure}[!t]
\centering
\includegraphics[width=\textwidth]{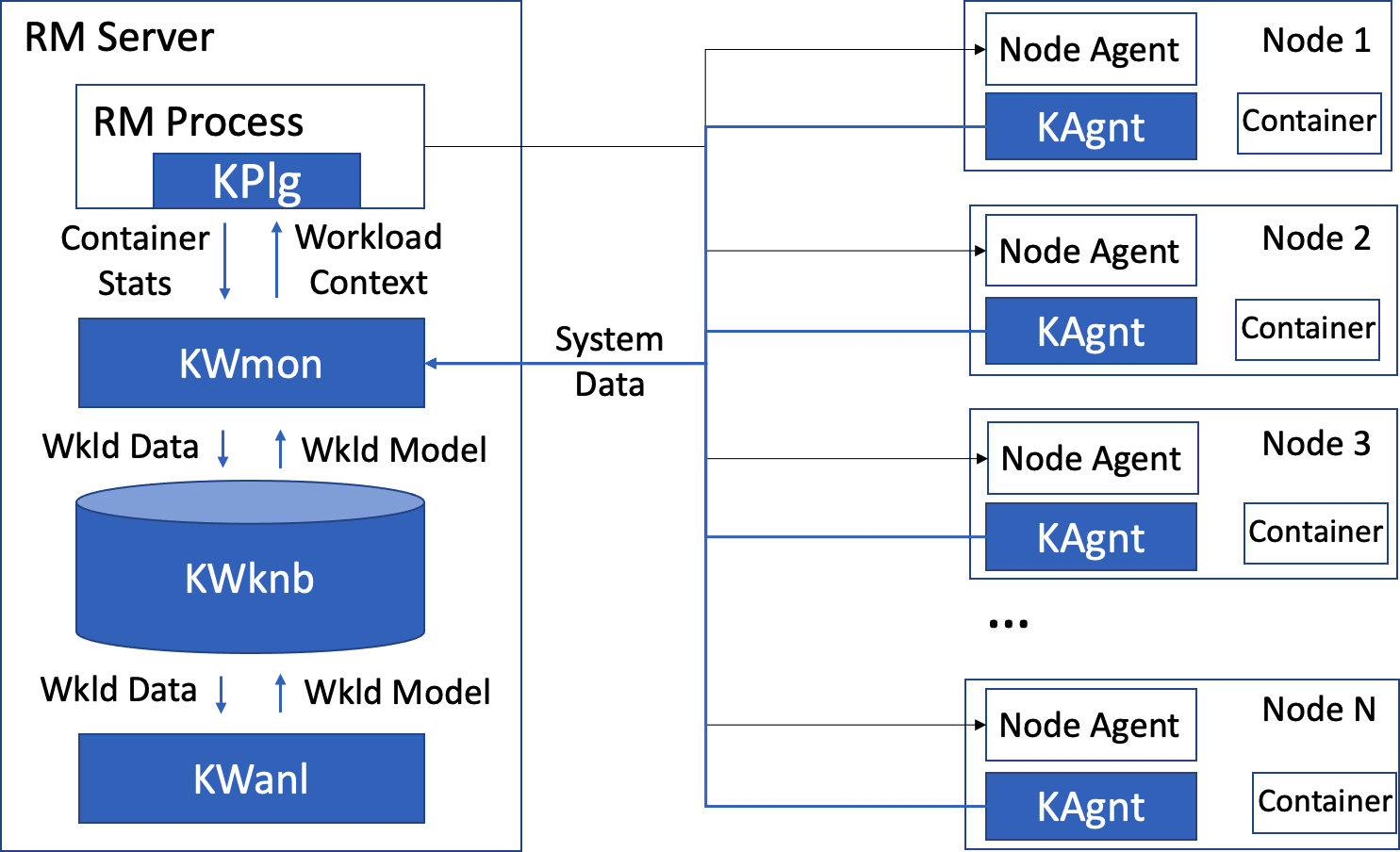}
\caption{The high-level KERMIT components shown in relation to the resource manager and the big data cluster.}
\label{kermit-components}
\end{figure}

Figure ~\ref{kermit-components} shows the high-level components of the KERMIT architecture, and how it relates to a typical big data cluster. The On-line sub-system interfaces with the Resource Manager process (RM Process in Figure ~\ref{kermit-components}) using a plug-in (KPlg). The core of the Workload Monitoring sub-system is the KERMIT Workload Monitor (KWmon) component. This component is a streaming engine that receives messages from the KPlg component and from the KERMIT system monitoring agents (KAgnt) deployed on cluster nodes.

The KERMIT Workload Analyser (KWanl in Figure ~\ref{kermit-components}) is the main component in the Off-line sub-system. The KWmon and the KWanl components implement real-time and batch machine learning pipelines form the core of the KERMIT architecture. The  details of each sub-system architecture are described in the sections below.

\subsection{Workload Classification, Prediction, and Optimization}

\begin{figure}[!t]
\centering
\includegraphics[width=\textwidth]{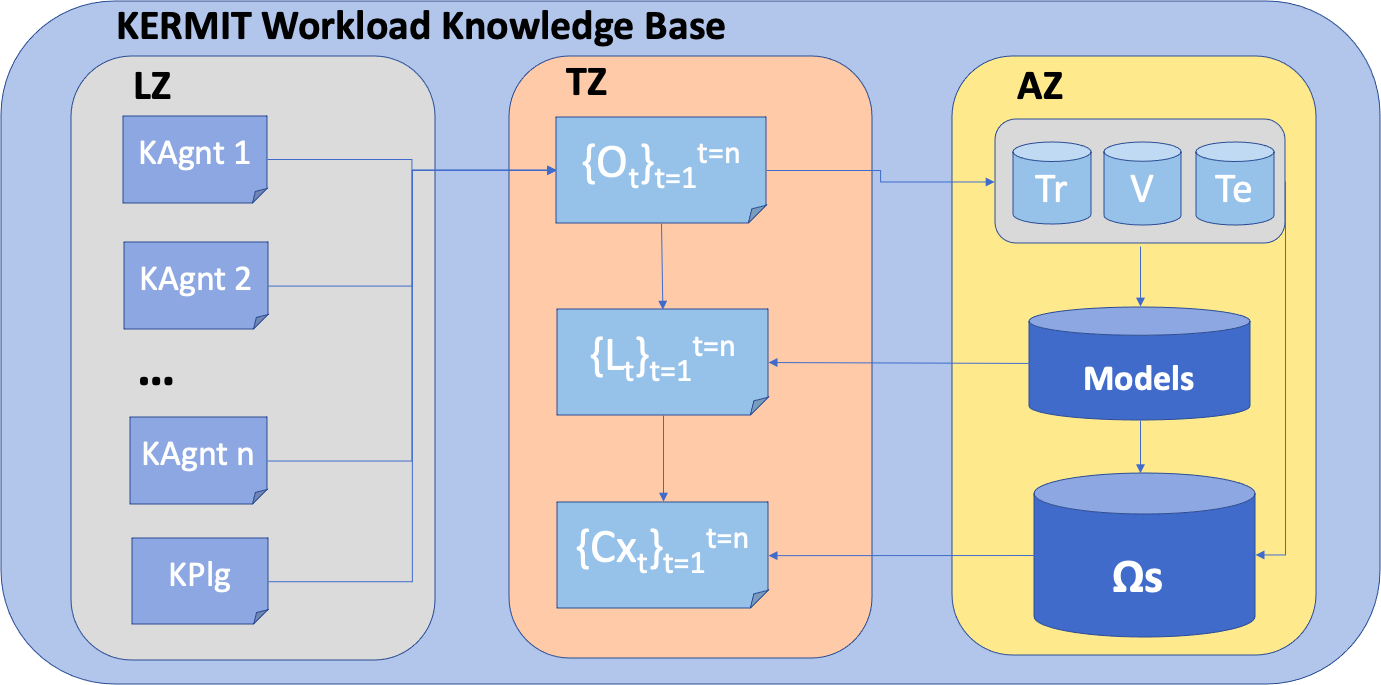}
\caption{Logical architecture of the KERMIT knowledge base.}
\label{kermit-knb-architecture}
\end{figure}

Figure ~\ref{kermit-knb-architecture} shows the logical architecture of the KERMIT workload knowledge base. The KERMIT workload knowledge base is implemented using a shared distributed file system, such as HDFS. The KERMIT workload knowledge base contains a Landing Zone (LZ in Figure ~\ref{kermit-knb-architecture}), a Transformation Zone (TZ), and an Analytics Zone (AZ). 

The raw time-stamped data generated by the KERMIT agents and the KERMIT plug-in components are stored in the LZ. These data are mostly loosely structured text files or log files. There is one file for each agent, and one for the KERMIT plug-in. The KERMIT workload monitor reads these data in real-time, treating each file as a streaming source, as new time-stamped data are appended. It transforms the time-stamped data into a structured format and aggregates them into observation windows $O_t$ with the associated feature vector ${\digamma}_t$. 

The KERMIT workload monitor applies the workload classification pipeline described in \cite{genkin2020workloadprediction} to transform the input stream of observation windows ${\{O_t}\}_{t=1}^n$ into a stream of labels ${\{Y_t\}}_{t=1}^n$, and writes out a sequence of workload context objects $\{\mathbb{C}_t\}_{t=1}^n$.

\begin{figure}[!t]
\centering
\includegraphics[width=\textwidth]{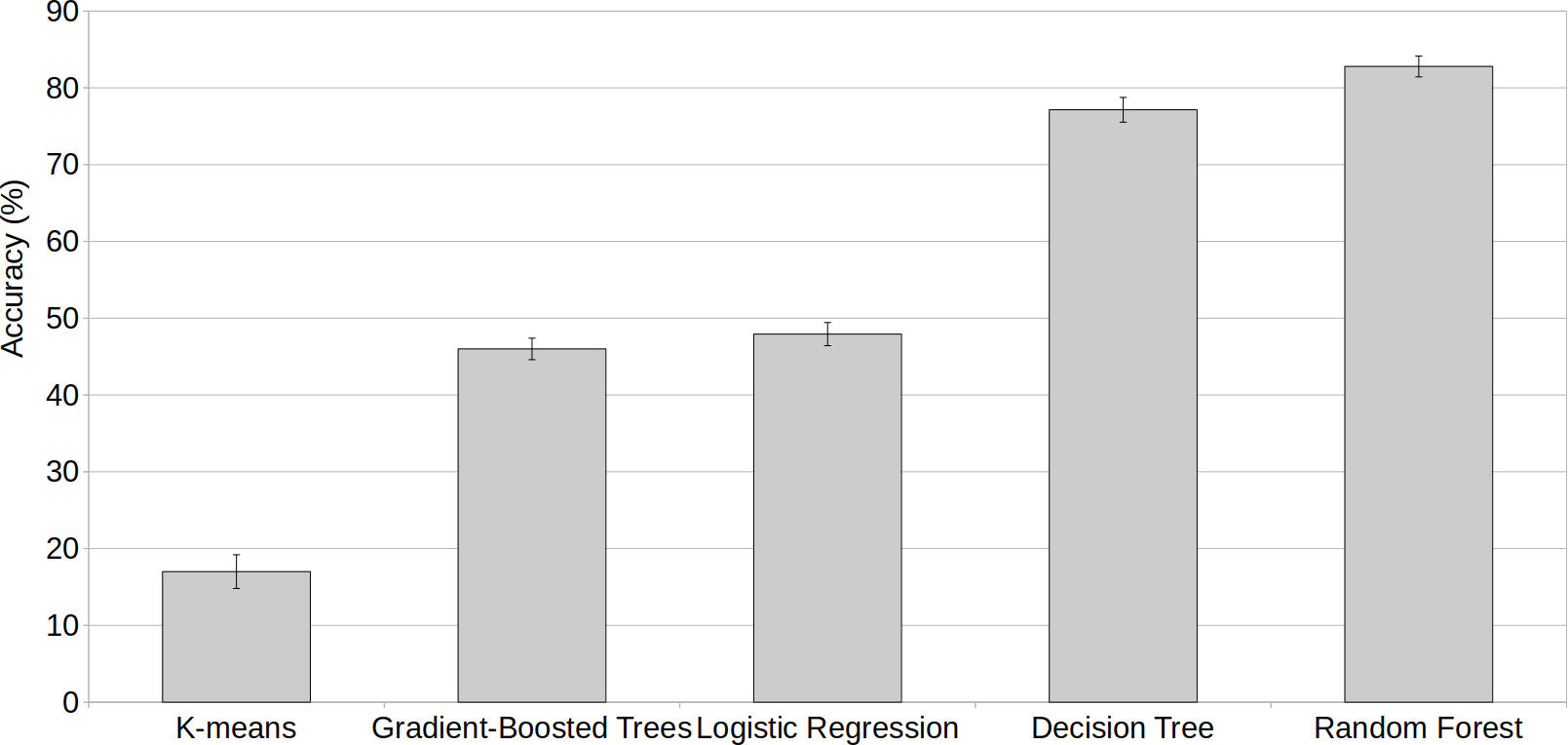}
\caption{The comparison of the classification accuracy for different machine learning algorithms \cite{genkin2019workloadclassification}.}
\label{classification-accuracy}
\end{figure}

\begin{figure}[!t]
\centering
\includegraphics[width=\textwidth]{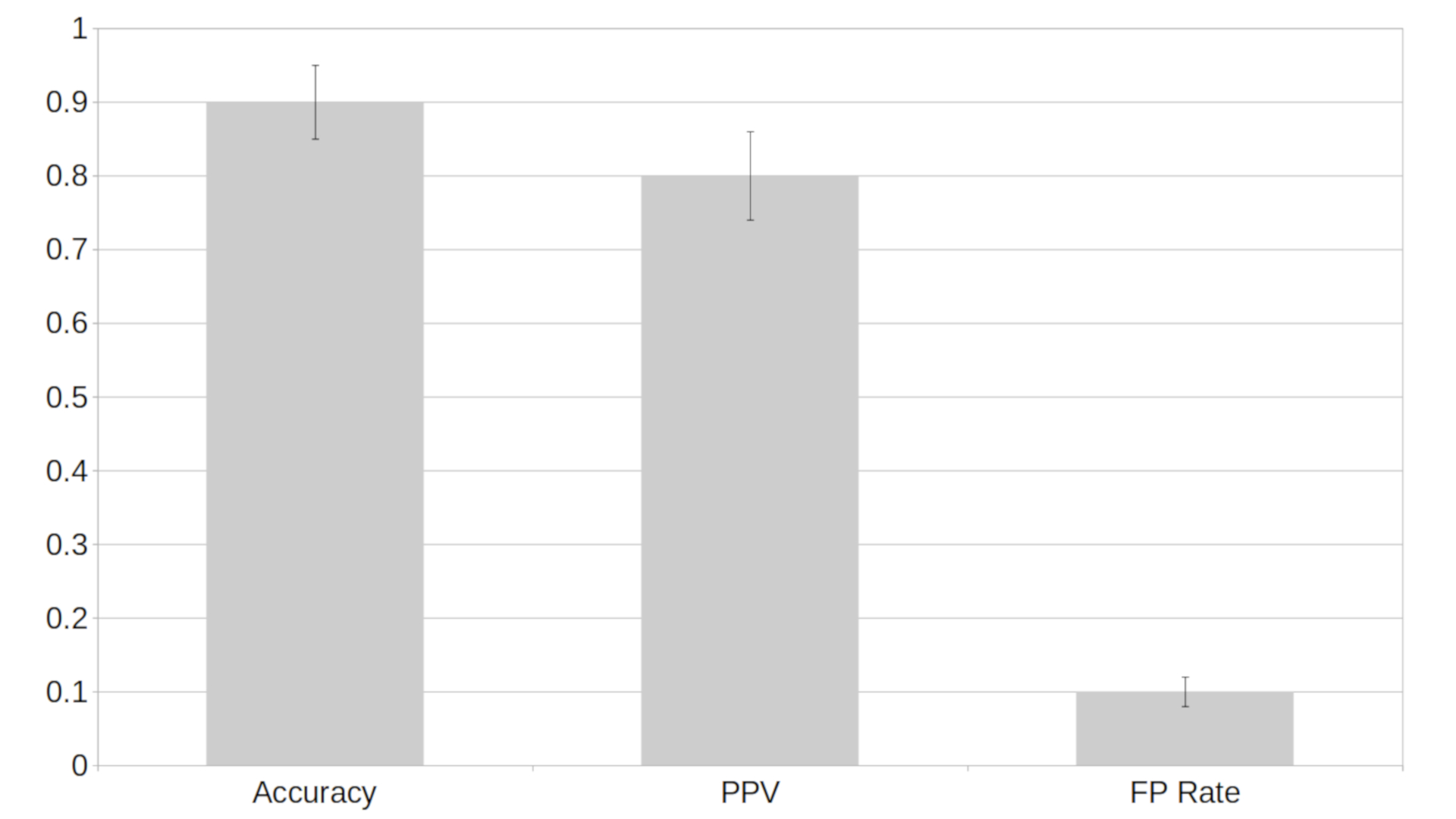}
\caption{TransitionClassfier algorithm performance \cite{genkin2020workloadprediction}.}
\label{transitionclassifier-summary}
\end{figure}

The workload context at observation window t, $\mathbb{C}_t$, contains the following information:

\begin{itemize}
\item The workload label for the current observation window t.
\item The predicted workload label for time horizon t+1.
\item The predicted workload label for time horizon t+5.
\item The predicted workload label for time horizon t+10.
\end{itemize}
 
The KERMIT plug-in code is called whenever the resource manager responds to a resource request from an analytic framework. The integration between the KERMIT plug-in and the resource manager is described in \cite{genkin2016kermit}. The KERMIT code intercepts the resource managers' response to the analytic frameworks' resource request and engages the low-overhead, conceptually simple, Explorer search algorithm  to find an optimal configuration \cite{genkin2016kermit}.

The Explorer demonstrated that it was capable of achieving up to 30\% better performance than rule-of-thumb tuning by a human practitioner, and up to 92.5\% tuning efficiency relative to the best possible tuning by a human practitioner. However, many real-world workloads are repetitive in nature. For example, the job to tally up the daily financial results is run at the same time every day. Some jobs are executed many times daily. If we consider that our definition of the term workload is more granular than a job - it stands to reason that the same workload type may be encountered many times during the day, or even the hour. 

It only makes sense to enhance the KERMIT plug-in architecture, described as the KERMIT Architecture in \cite{genkin2016kermit}, with the ability to recognize workloads that have already been executed, and for which the optimal configuration has already been found. This allows KERMIT to avoid repeating the same parameter search multiple times, and achieve further performance gains under realistic workload conditions.

The KERMIT plug-in extends the KERMIT Analyser component presented in \cite{genkin2016kermit} with the capability to read the workload context stream generated by the KERMIT workload monitor. When called by resource manager it first checks the workload monitor stream output and retrieves the workload context object. It then checks the label of the currently executing workload, and retrieves the workload descriptor object from KERMIT workload knowledge base.

The workload descriptor object contains the following items of information:

\begin{itemize}
\item Statistics for each feature in the workload feature vector (described further below).
\item Centroid values for the workload (described further below).
\item A true of false flag indicating whether or not an optimal configuration has been found.
\item A a set of configuration values to be used. 
\end{itemize}

The main high-level algorithm used by the KERMIT plug-in is described in Algorithm ~\ref{kplg-algorithm}. When the resource manager calls the KERMIT plug-in code in response to a resource request from one of the analytic frameworks, the Analyser component reads the workload context stream $\{\mathbb{C}_t\}_{t=1}^n$. It reads in the latest context $\mathbb{C}_t$, and compares the current time with the observation window associated with the context to make sure that the plug-in and the KERMIT workload monitor are in-sync. If they are not then an error is logged and a default configuration is used until the error is resolved.

Once the context has been read in, the plug-in checks the workload type label for the current observation window $\mathbb{C}_t.currentLabel$. When the KERMIT workload monitor first starts and the different workload types have not yet been determined, the type will be UNKNOWN. In this case the KERMIT plug-in will simply use the default configuration $ \mathfrak{J}^D$ as the optimal configuration $\mathfrak{J}_i^o$ for this workload. The plug-in will wait until the off-line sub-system workload discovery catches up, and will continue to check the current workload type each time it is called by the resource manager.

Once the KERMIT plug-in encounters a current workload label that is known, it will check the KERMIT WorkloadDB to see if an optimal configuration for this label has already been established. If so, it will simply retrieve this configuration from the WorkloadDB.

If there is no optimal configuration associated with this workload, then the plug-in will check if workload drift has been detected. If so, then there will be a configuration in the WorkloadDB associated with this workload label, but this configuration will not be optimal. In this case the KERMIT plug-in will retrieve this configuration and pass it to the Explorer algorithm to initiate a local search described in \cite{genkin2016kermit}. Once the local search finds the optimal configuration, the plug-in will update the KERMIT WorkloadDB with this configuration, and set the optimal configuration field in the database to the value "True".

If workload drift has not been detected then the WorkloadDB will not have a sub-optimal configuration stored for the workload because the workload has just been detected by the off-line sub-system. In this case the KERMIT plug-in will start the Explorer algorithm's global search described in \cite{genkin2016kermit}. Once the global search finds the optimal configuration the KERMIT plug-in will update the WorkloadDB with this configuration, and set the optimal configuration field in the database to the value "True".

\begin{algorithm}
\caption{Main KERMIT plug-in algorithm.}
\begin{algorithmic}
\REQUIRE $\{\mathbb{C}_t\}_{t=1}^n \neq \{\}$ \COMMENT{Workload context stream must be started and available.}
\REQUIRE $O_t \neq \emptyset$ \COMMENT{Current observation window.}
\ENSURE $\mathfrak{J}_i^o$ \COMMENT{The optimal configuration for this workload}.

\IF{$\mathbb{C}_t.currentLabel$ is UNKNOWN}
	\STATE $\mathfrak{J}_i^o \leftarrow \mathfrak{J}^D$ \COMMENT{Use the default configuration for unknown workloads.}
\ELSE
	\IF{$\mathbb{C}_t.currentLabel$ has optimal config in WorkloadDB}
		\STATE $\mathfrak{J}_i^o \leftarrow$ get optimal config from WorkloadDB
	\ELSE
		\IF{$\mathbb{C}_t.currentLabel$ has workload drift}
			\STATE $\mathfrak{J}_i^o \leftarrow$ Explorer.localSearch($\mathfrak{J}_i$) \COMMENT{Do a local search starting with the last good configuration.}
		\ELSE			
			\STATE $\mathfrak{J}_i^o \leftarrow$ Explorer.globalSearch()
		\ENDIF
		\STATE Update WorkloadDB with $\mathfrak{J}_i^o$
	\ENDIF
\ENDIF

\end{algorithmic}
\label{kplg-algorithm}
\end{algorithm}

The algorithms used to identify different workload types and detect workload drift are discussed in the section below.

\section{The KERMIT Off-Line Sub-System Architecture}

\begin{figure}[!t]
\centering
\includegraphics[width=\textwidth]{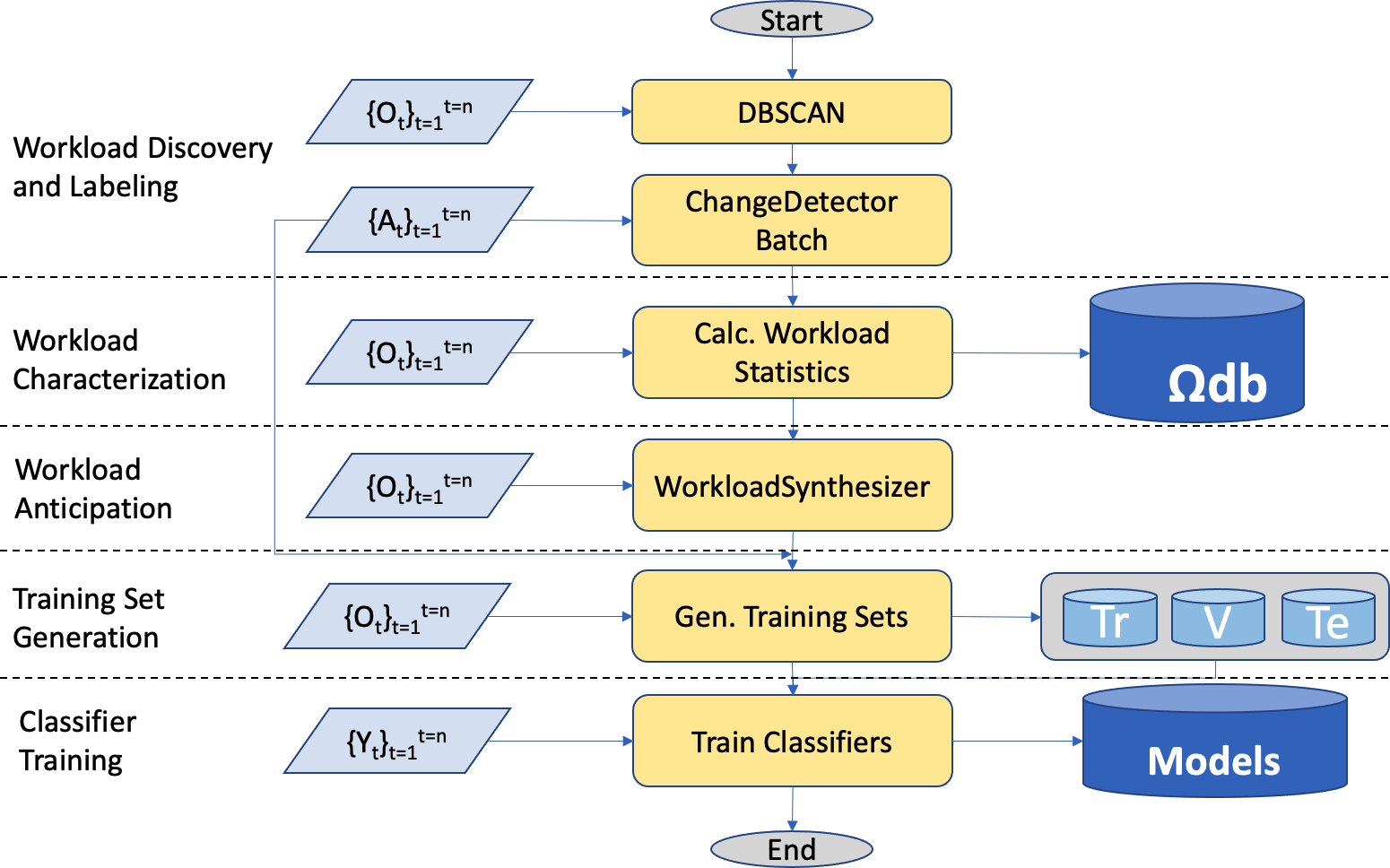}
\caption{The high-level steps in the off-line sub-system processing pipeline.}
\label{off-line-pipeline}
\end{figure}

Figure ~\ref{off-line-pipeline} shows the high-level processing pipeline flow in the off-line sub-system. The high-level component KERMIT workload analyser (KWanl in Figure ~\ref{kermit-components}) implements the off-line analytic pipeline which performs the following stages of processing:

\begin{enumerate}
 \item Workload discovery and labeling.
 \item Workload characterization.
 \item Workload anticipation.
 \item Training set generation.
 \item Classifier training. 
 \end{enumerate} 

Sections below begin by discussing processing pipeline and the algorithm used to detect and identify new workloads. The next section builds on this discussion to describe the algorithm for detecting workload drift.

\subsection{Workload Discovery, Characterization, and Drift Detection} \label{ch-autonomic-arch-offline-discovery}

As discussed in the sections above, the KERMIT workload monitor stores the aggregated stream of workload windows ${\{O_t}\}_{t=1}^n$ in the transformation zone of the KERMIT knowledge base (see Figure ~\ref{kermit-knb-architecture}). The high-level algorithm for workload discovery, characterization and drift detection is given in Algorithm ~\ref{discovery-and-drift-algorithm}.

\begin{algorithm}
\caption{The workload discovery and drift detection algorithm.}
\begin{algorithmic}
\REQUIRE ${\{O_t}\}_{t=1}^n \neq \emptyset$ \COMMENT{Landed observation window time-series.}
\ENSURE $\mathfrak{\{Y\}_j^k}$ \COMMENT{Set of identified workload labels.}
\STATE run ChangeDetector.batch() to identify transition windows
\STATE extract transition windows from ${\{O_t}\}_{t=1}^n$
\STATE run DBSCAN on ${\{O_t}\}_{t=1}^n$ to get a set of clusters 
\FORALL{clusters in the set}
	\STATE calculate workload characterization statistics 
	\IF{find match in WorkloadDB is True}
		\STATE $Y_j$ gets matched label from WorkloadDB 
		\IF{L2 norm between the mean vectors of workload characterisations differ by more than $\epsilon$}
			\STATE update isDrifting to True in WorkloadDB
			\STATE update workload characterisation for matching label in WorkloadDB with new data
		\ENDIF
	\ELSE
		\STATE generate new label for the new workload
		\STATE insert new workload label and characterization data into WorkloadDB
	\ENDIF	
\ENDFOR

\end{algorithmic}
\label{discovery-and-drift-algorithm}
\end{algorithm}

The algorithm begins by using the ChangeDetector component in batch mode to scan the persisted time series of observation window data, and flag workload transition windows as described in \cite{genkin2020workloadprediction}. The logic of the batch operation is exactly the same as in the real-time use case. The workload transition windows are then removed from the original set into a separate set, and clustering analysis is performed on the now filtered set of workload observation windows.

\begin{figure}[!t]
\centering
\includegraphics[width=\textwidth]{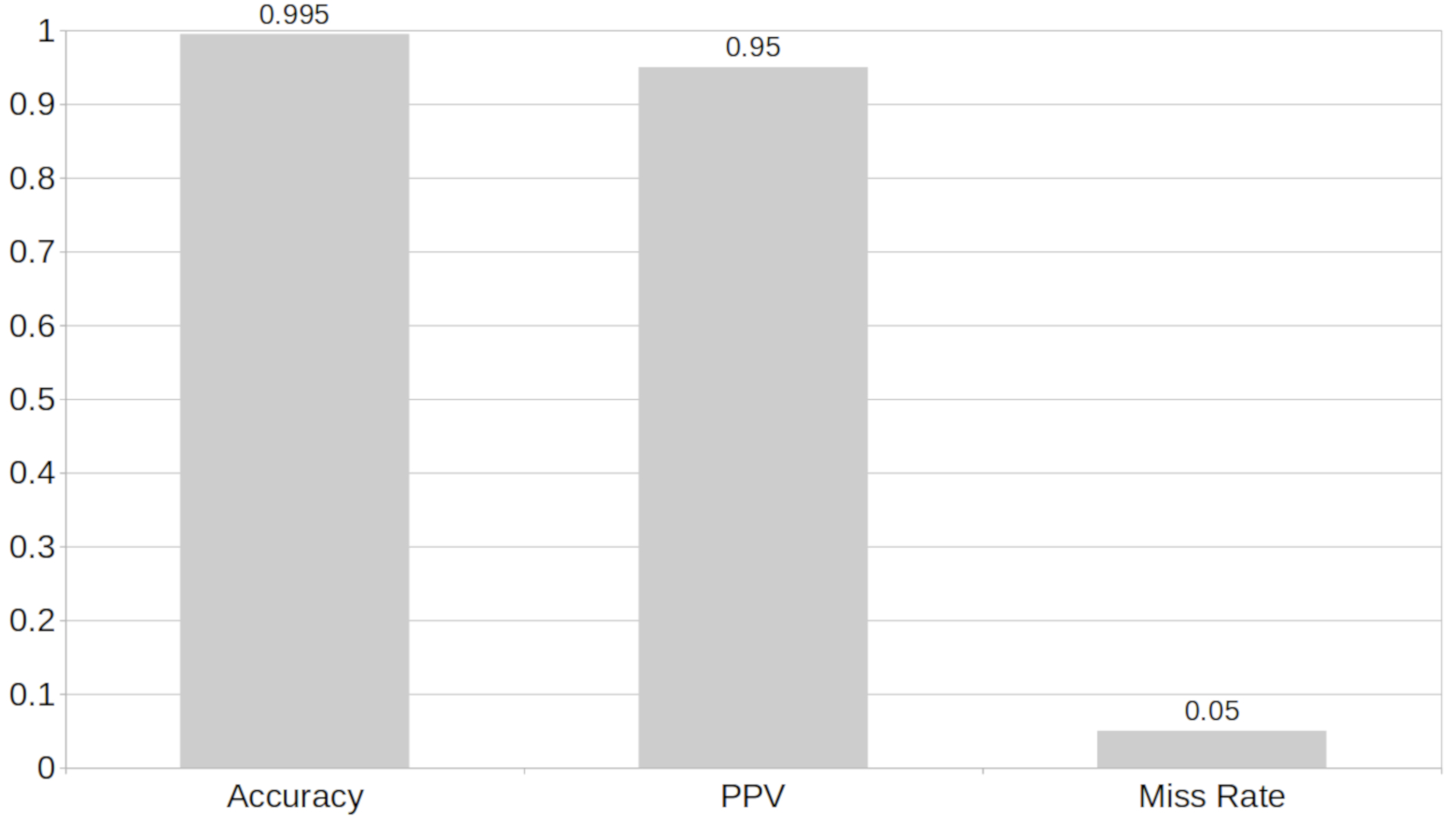}
\caption{ChangeDetector algorithm performance \cite{genkin2020workloadprediction}.}
\label{changedetector-summary}
\end{figure}

\begin{figure}[!t]
\centering
\includegraphics[width=\textwidth]{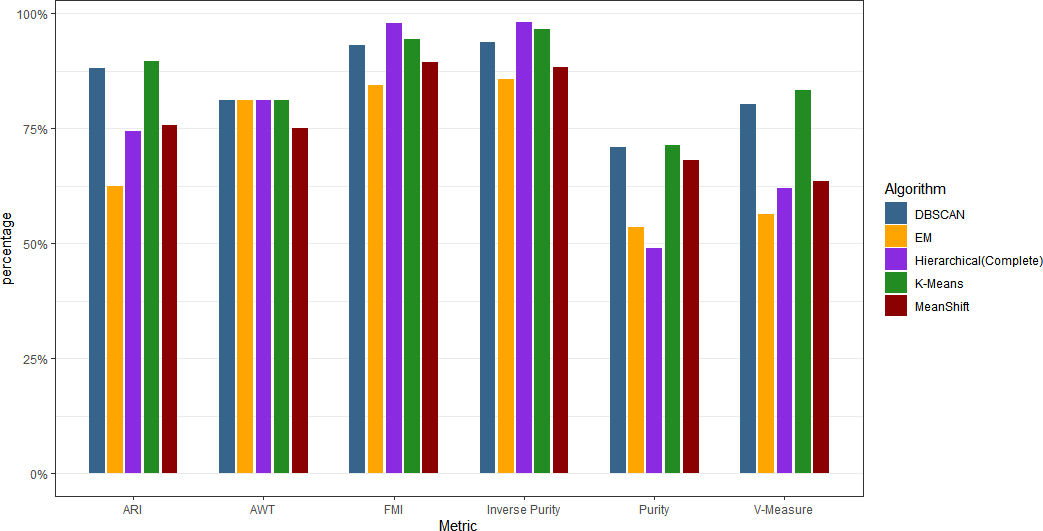}
\caption{Workload discovery performance for clustering algorithms.}
\label{clustering-algo-comparison}
\end{figure}

Figure ~\ref{clustering-algo-comparison} shows the key performance metrics for several different clustering algorithms. The effectiveness of each algorithm was evaluated using time-series workload data recorded during the execution of Apache Hadoop and Spark benchmarks. Clustering results were compared to ground truth interpretation made by a human specialist using Apache Hadoop and Spark logs. 

The key metrics indicated in the figure are Awt and Purity. Purity indicates how many of the observation windows were classified correctly by the on-line sub-system. The Awt metric is also an accuracy-type metric. It measures how accurately the algorithm was able to identify different workload types. For example, if the benchmark executed 3 different workload types and the algorithm detected 3 clusters whose centroids call within the observation window range of each workload type, then the Awt metric for this algorithm would be 100\%.

Workload discovery in KERMIT is accomplished by running the DBSCAN algorithm on filtered observation window data.  DBSCAN identifies clusters within the observation window data. Each cluster represents a distinct workload type. 

The next step (see Algorithm ~\ref{discovery-and-drift-algorithm}) is to check if the newly identified workloads have been encountered before by calculating the workload characterization statistics for each cluster, and comparing with workload characterization statistics for workloads already identified and stored in WorkloadDB. This is accomplished by using the ChangeDetector off-line to compare statistical data.

Workload characterization involves calculating the relevant statistics for each subset of observation windows that were grouped by the DBSCAN algorithm into the corresponding cluster. A full set of statistics, including the mean, the standard deviation, the max, the min, the 90th percentile, and the 75th percentile are calculated. This set of statistics is the workload characterization.

If a matching workload characterization is found in WorkloadDB (as identified by the ChangeDetector), then this is an existing workload. All of the windows in the cluster get tagged with the matched workload label from WorkloadDB.

The next step is to check the workload for drift. This is accomplished by calculating the L2 norm of the distance between the mean vectors of the new cluster and the one stored in WorkloadDB. If the difference is larger than the configurable hyper-parameter $\epsilon$, then drift has occured, and the WorkloadDB is updated with the workload characterization for the new cluster.

If no matching workload is found in the WorkloadDB, then the KERMIT workload analyser generates a unique integer label for the cluster. The generated workload labels do not need to be human-legible. They just need to be unique to each identified cluster of observation windows. Currently KERMIT implements a simple integer counter, because this facilitates the generation of libsvm files for model training. Then new label, along with the workload characterization, is inserted into WorkloadDB.  

\begin{figure}[!t]
\centering
\includegraphics[width=\textwidth]{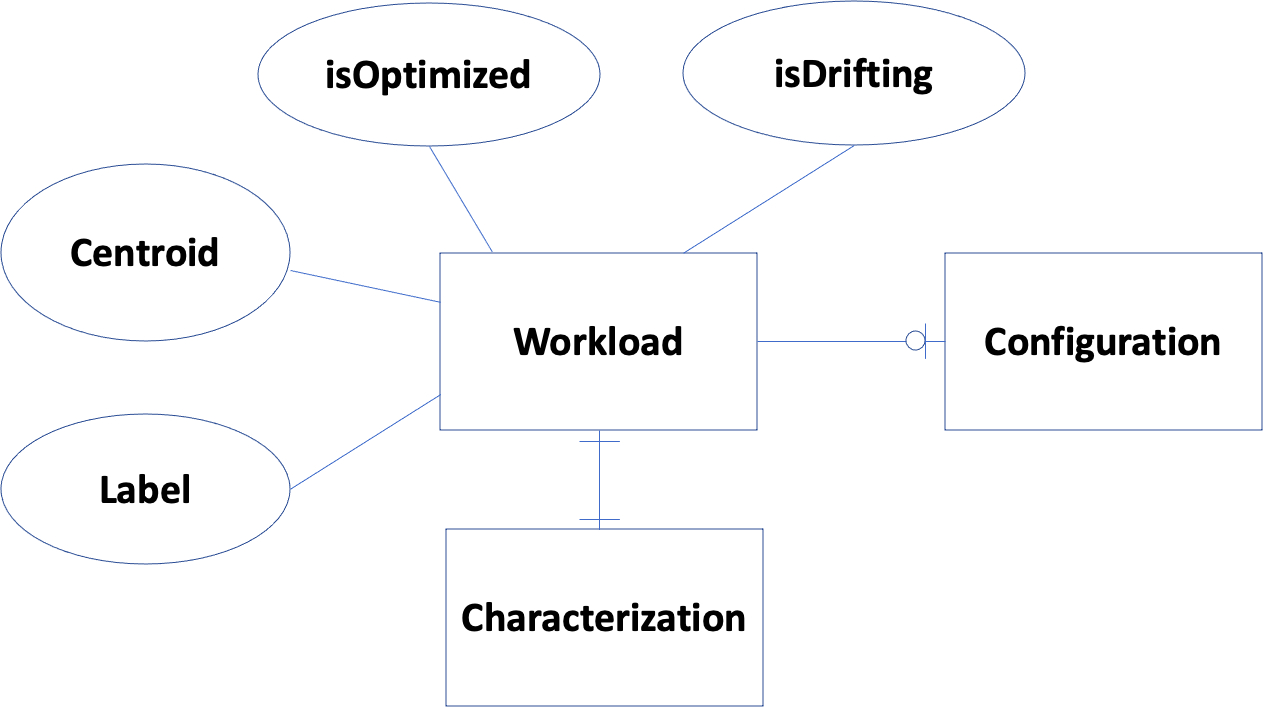}
\caption{Entity-relationship model of the WorkloadDB schema.}
\label{workloaddb-er-model}
\end{figure}

Figure ~\ref{workloaddb-er-model} shows the WorkloadDB data model. Each workload is uniquely identified by its unique automatically generated label. Each workload contains the workload characterization statistics, a true/false field indicating whether the optimal configuration has been found, and a true/false field indicating whether workload drift has been detected. Each workload can have one configuration stored in the WorkloadDB. This configuration may or may not be the optimal configuration.

When a workload is initially identified, it will not have a configuration associated with it. This is because the configuration search is performed in real time by the on-line sub-system. Once the KERMIT plug-in performs the global search for the workload, it will update the WorkloadDB with the optimal configuration, and set the field indicating whether the optimal configuration has been found to True.

The next time clustering analysis is performed this occurs on a set interval, and on a new set of ${\{O_t}\}_{t=n+1}^{n+k}$ data collected during the last interval, where $k$ is a constant hyper-parameter that controls the length or the batch used for clustering analysis. The entire process described in Algorithm ~\ref{discovery-and-drift-algorithm} is repeated.

This approach allows KERMIT to continuously learn new workloads. New workloads are added to the WorkloadDB as they are detected during the off-line batch processing. Already known workloads are protected against workload drift because their characterizations are regularly updated. Workloads are never deleted from WorkloadDB. Thus KERMIT retains a long-term memory of workloads, and the ability of the KERMIT on-line sub-system to recognize workloads improves over time.

\subsection{Automated Classifier Training}

The KERMIT on-line analytic pipeline includes several classifiers described in \cite{genkin2019workloadclassification}, and \cite{genkin2020workloadprediction}. Paragraphs below provide a quick overview of their function and purpose, discuss their training requirements and describe how the process is automated.

The KERMIT on-line classification pipeline uses the following classifiers:

\begin{itemize}
\item \textit{ChangeDetector}. This statistical classifier is a binary classifier that simply uses the Welch's statistical test to distinguish steady state processing from workload transitions. This classifier does not require off-line training.
\item \textit{WorkloadClassifier}. This classifier is based on the random forest ensemble algorithm This is a supervised classifier that does require off-line training. 
\item \textit{TransitionClassifier}. This classifier is also based on the random forest ensemble algorithm. It also is a supervised classifier that does require off-line training.
\item \textit{ZSL Workload  Classifier}. This component, described in \cite{genkin2020zslmultiuser} re-uses the \textit{WorkloadClassifier} class, and introduces the \textit{WorkloadSynthesizer} component. This component needs to be trained off-line. It also generates synthetic class instances, which need to be merged into the training process for the \textit{WorkloadClassifier}. 
\item \textit{WorkloadPredictor}. This component is based on an LSTM neural network algorithm.
\end{itemize}

The training pipeline performs the following high-level steps (some of the steps performed as part of workload discovery are repeated for completeness):

\begin{enumerate}
\item Extract observation window range for each workload in WorkloadDB.
\item Use workload observation window id set to extract a set of analytic windows from the analytic window stream ${\{A_t}\}_{t=1}^n$ created by the Workload Monitor. For every observation window there is a matching analytic window. This becomes the training set for the WorkloadClassifier $Dp_{\Omega}^{Tr}$.
\item Establish window ranges for workload transitions by scanning the sequence of analytic windows and marking ranges of windows that connect window sets that belong to each workload cluster, and correlating with workload transitions identified during the workload discovery phase.
\item Generate labels for each workload transition type. The same algorithm is used for label generation as that used for workloads. The labels don't need to be human-readable, just unique and consistent.
\item Transform the analytic window sequence ${\{A_t}\}_{t=1}^n$ to a rate of change sequence ${\{A'_t}\}_{t=1}^n$.
\item Extract transition windows from the transformed sequence - this forms the training set for the TransitionClassifier $D_{\Delta}^{Tr}$.
\item Execute the WorkloadSyntesizer component on $Dp_{\Omega}^{Tr}$ to account for possible anticipated hybrid, multi-user workloads. This involves the following steps: 
	\begin{enumerate}
 		\item Generating the Class Descriptor file described in \cite{genkin2020zslmultiuser}. Each workload entry in the Workload DB is used as a pure class. Possible hybrid workloads are constructed by pairing the pure workloads.
 		\item Generate labels for the anticipated, hybrid workloads using the same algorithm as for pure workload classes.
 		\item Update the WorkloadDB with synthetic class prototypes - these contain the same information as the workload characterizations calculated for the seen classes.
 		\item Merge the synthetic workload instances  with the observed workload instances to construct the final, merged, WorkloadClassifier training set $D_{\Omega}^{Tr}$.
 	\end{enumerate}
\item Generate the training set for the WorkloadPredictor component $D_{\mho}^{Tr}$ by extracting segments from the label sequence ${\{Y_t}\}_{t=n+1}^{n+k}$.
\item Train the classifiers.
\end{enumerate}

Most of the steps described above, with the exception of steps needed to train the WorkloadClassifier, can be executed in parallel given sufficient compute resources.

\section{Conclusion}

This paper presents the first architecture intended for autonomic optimization of big data workloads. The KERMIT architecture implements an autonomic feedback look that includes on-line and off-line processing stages. It uses machine learning pervasively to analyze workload characteristics, identify new workload types, detect changes in real-time, classify workloads, and predict future workload types and characteristics. Change detection, workload classification, workload prediction, and parameter search are performed on-line, in real time. Classifier training is performed off-line as a batch machine learning pipe-line.

Experimental investigations focused on the critical proof points of the autonomic feedback loop demonstrate that the KERMIT architecture can:

\begin{itemize}
\item Real-time workload classification with up to 90\% accuracy \cite{genkin2020workloadprediction}. 
\item Detect workload changes in real-time with up to 99\% accuracy \cite{genkin2020workloadprediction}.
\item Predict workload type with up to 96\% accuracy \cite{genkin2020workloadprediction}.
\item Anticipate new, unseen workload types, and classify them with up to 83\% accuracy \cite{genkin2020zslmultiuser}.
\end{itemize}

The KERMIT architecture, as discussed above, can anticipate the appearance of new unseen, multi-user, hybrid workloads that can present a mix of characteristics observed with the currently identified workloads. This is a capability that has not been previously described in any of the earlier works focused on small data and cloud workload segments.   

The ability to adjust to workload drift is another key element of the KERMIT architecture. The on-line sub-system will do its best to classify workloads in real-time. If a previously unseen and unanticipated workload is encountered, the KERMIT on-line sub-system will initially classify it as one of the known workload types, with the closest characteristics, and use the best available configuration for that workload. This is often better, in terms of reducing the tuning overhead, then immediately performing a global search for that workload.The new workload will be discovered by the off-line sub-system the next time it performs clustering analysis.  

This architecture can operate with minimal configuration by a human administrator. Although there are still a number hyper-parameters that need to be set, these, unlike many of the Apache Hadoop and Spark configuration settings, do not require frequent tuning. For the most part these hyper-parameters can be left at their default settings. These default settings (for example the $\mu$ hyper-parameter for the DBSCAN algorithm) in many cases apply to a broad range of conditions and are well-documented in the scientific and technical literature.

\bibliography{autonomic-architecture-for-big-data}
\bibliographystyle{unsrt}
\end{document}